\documentclass[aps,prl,twocolumn,superscriptaddress,showpacs]{revtex4}

\usepackage{graphicx}
\usepackage{tabularx}
\begin{document}

\title{Non-parabolicity of the conduction band of wurtzite GaN}

\author{S. Syed}
\affiliation{Department of Applied Physics and Applied
Mathematics, Columbia University, New York, New York 10027}
\author{J. B. H\'{e}roux}
\affiliation{Department of Electrical Engineering, Columbia
University, New York, New York 10027}
\author{Y. J. Wang}
\affiliation{National High Magnetic Field Laboratory, Florida
State University, Tallahassee, FL 32306}
\author{M. J. Manfra}
\affiliation{Bell Laboratories, Lucent Technologies, Murray Hill,
NJ 07974}
\author{R. J. Molnar}
\affiliation{MIT Lincoln Laboratory, Lexington, MA 02420-0122}
\author{H. L. Stormer}\affiliation{Department of Applied Physics and Applied
Mathematics, Columbia University, New York, New York
10027}\affiliation{Bell Laboratories, Lucent Technologies, Murray
Hill, NJ 07974}\affiliation{Department of Physics, Columbia
University, New York, New York 10027}

\begin{abstract}
Using cyclotron resonance, we measure the effective mass, $m$*, of
electrons in AlGaN/GaN heterostructures with densities,
$n_{2D}\sim 1-6\times10^{12}$cm$^{-2}$. From our extensive data,
we extrapolate a band edge mass of $(0.208\pm0.002) m_e$. By
comparing our $m$* data with the results of a multi-band
\textbf{k.p} calculation we infer that the effect of remote bands
is essential in explaining the observed conduction band
non-parabolicity (NP). Our calculation of polaron mass corrections
-- including finite width and screening - suggests those to be
negligible. It implies that the behavior of $m$*$(n_{2D})$ can be
understood solely in terms of NP. Finally, using our NP and
polaron corrections, we are able to reduce the large scatter in
the published band edge mass values.
\end{abstract}
\pacs{73. 20.At, 76. 40.+b, 73. 40.-c} \maketitle

The magnitude of the conduction band non-parabolicity (NP) in
wurtzite GaN currently remains controversial. NP of a band can be
probed by measuring the carrier effective mass $m$* as a function
of energy. Such experiments have been performed in the past in
both bulk and in two-dimensional electron systems (2DES). The
deduced band edge mass values, $m_0$*, however, exhibit
considerable scatter. Using cyclotron resonance (CR), Drechsler
\textit{et al.} \cite{Drechsler95} determined $m_0$* in bulk
wurtzite GaN to be 0.20$m_e$, where $m_e$ is the free electron
mass. Other methods such as infrared reflectivity on electron
plasma \cite{Perlin96} and spectroscopy on shallow donors
\cite{Moore95,Witowskia99,Meyer95} in bulk GaN have yielded
0.220$m_e < m_0^* <$ 0.236$m_e$. An even wider range of values for
the band edge mass, 0.185$m_e < m_0^* <$ 0.231$m_e$, emerge from
experiments in AlGaN/GaN heterostructures. From the temperature
dependence of Shubnikov-de Haas (SdH) oscillations of 2DES, Lin
\textit{et al.} deduced $m_0$* = 0.22$m_e$ \cite{Lin98} while Hang
\textit{et al.} reported $m_0$* = 0.185$m_e$ \cite{Hang01}.
Cyclotron resonance experiments on heterostructures have revealed
0.223$m_e < m_0^* < $ 0.231$m_e$ \cite{Knap96,Knap97}. This spread
in $m_0$* suggests that the ``extrapolation'' from the various
experiment values $m$*$(E)$ to the band edge remains poorly
controlled.

Using CR we have measured $m$* in a series of high mobility
($\sim$20,000 cm$^2$/Vsec) AlGaN/GaN structures. Our
heterostructures are all grown by molecular beam epitaxy (MBE) on
GaN templates prepared by hydride vapor phase epitaxy (HVPE). The
specimens are described in detail elsewhere
\cite{CRsplit03,Manfra02}. Our data cover a density range of
$1-6\times10^{12}$ cm$^{-2}$. In these 2D systems this implies
energies from $\sim$27meV to $\sim$120meV above the band edge due
to electron confinement and band filling. Therefore, our mass data
probe the NP of the GaN conduction band in small steps over a wide
energy range. We also perform extensive \textbf{k.p} calculations
and determine that instead of the commonly used two-band model, a
multi-band model is required to explain our experimental results.
Additionally, our calculations on polaron correction of the
effective mass of 2D electrons in GaN show them to be at the 1$\%$
level, considerably less than previously thought \cite{Knap96}. A
detailed comparison between our data and \textbf{k.p} calculation
sets the band mass value to $m_0^* =(0.208\pm0.002) m_e$ and, when
applied to other investigators' results, considerably reduces the
spread in band edge mass values.

A Fourier transform spectrometer with light pipe optics and a
composite Si bolometer was used for the detection of far-infrared
transmission. Magnetic field was applied normal to the 2D electron
layer. The carrier density of each sample was determined
\textit{in situ} from the Shubnikov-de Haas (SdH) oscillations of
the 2DES. All CR and SdH experiments were conducted at 4.2K.

\begin{figure}
\includegraphics{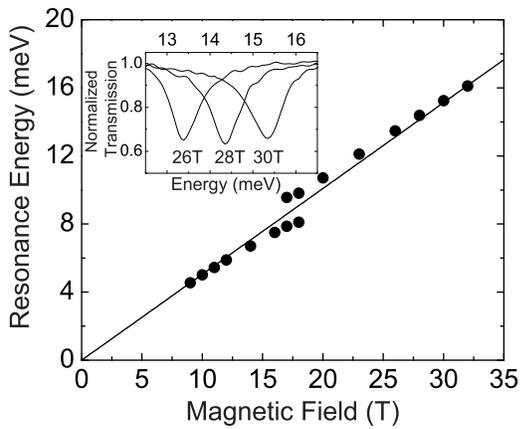}
\caption{\label{Fig.1} Resonance energies vs. $B$ of a sample with
$n_{2D}=2.3\times10^{12}$ cm$^{-2}$. High and low field resonances
can be fit with a single straight line, giving an effective mass
of 0.228$m_e$. Near 18T a level anti-crossing results in a
splitting of the CR (see text). Inset: Transmission data for
$B$=26, 28, and 30T, normalized to the spectrum at $B$=0T.}
\end{figure}

Fig.\ref{Fig.1}  shows the cyclotron resonance energies versus
magnetic field, \textit{B} of a sample with
$n_{2D}=2.3\times10^{12}$ cm$^{-2}$. All data are taken with a
resolution of 0.24 meV. The inset shows high field transmission
spectra normalized to the spectrum taken at \textit{B}=0T. The
solid line in Fig.1 is a fit to the data at high fields,
\textit{B}$>$27T, and low fields, \textit{B}$<$12T, resulting in
an effective mass of $m$* = 0.228$m_e$. For
15T$<$\textit{B}$<$25T, there is a pronounced deviation from this
straight line. This anomaly in the CR represents a recent
discovery, which is being analyzed and published elsewhere
\cite{CRsplit03}. Here, we observe that this anomaly is limited to
a finite field region outside which all CR data can be fit by a
straight line.

\begin{figure}
\includegraphics{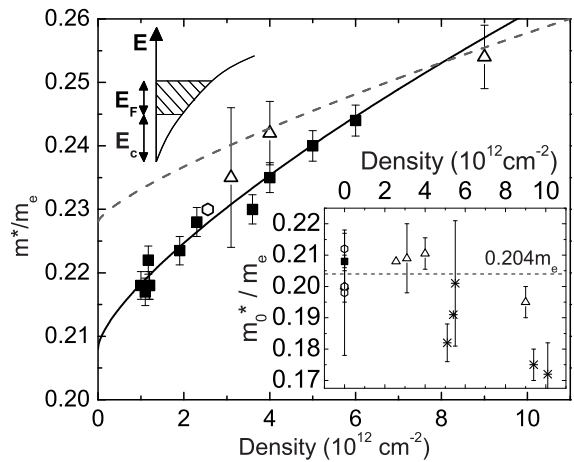}
\caption{\label{Fig.2} Effective mass, $m$*, vs. 2D electron
density. Our data are shown as solid squares. Data from Ref.
\cite{Knap96,Knap97} and \cite{Wang96} are shown as open symbols.
The dashed line is a two band fit to the results of Ref.
\cite{Knap97} according to Eq.\ref{eq.1} with $K$=1. The solid
line represents a fit with $K$=2.5, which accounts for the
influence of additional, higher conduction bands. Inset: Values of
the band edge mass from Refs.
\cite{Drechsler95,Knap96,Knap97,Hang01,Lin98,Perlin96,Moore95,Witowskia99,
Meyer95,Wang96} and this work, after NP and polaron corrections.
CR results are shown as open triangles, donor spectroscopy data as
open hexagons and SdH data as stars. Our result of $m_0^* =
(0.208\pm0.002)m_e$ is shown as a solid square. Averaging all data
(except SdH) yields $m_0^*$ = 0.204 $m_e$ illustrated by a dashed
line (see text). }
\end{figure}

We have measured the effective mass in eleven samples with carrier
density, $n_{2D}$, ranging from $1-6\times10^{12}$ cm$^{-2}$. In
all cases, we observed either a broadening or a splitting of the
CR line at intermediate fields but could fit the data away from
this regime as well as the data seen in Fig. \ref{Fig.1}.
Fig.\ref{Fig.2} shows the dependence of $m$* on $n_{2D}$. For
comparison, we also plot data from Ref.
\cite{Knap96,Knap97,Wang96}. The mass data from the different
references are located in the general vicinity of our results but,
due to their considerable error bar or sparsity, are difficult to
extrapolate to zero density. The combined data of Fig.\ref{Fig.2}
show an increase in $m$* by $\sim$17$\%$ as $n_{2D}$ changes from
$1-9\times10^{12}$ cm$^{-2}$. The rise in $m$* with $n_{2D}$
reflects the non-parabolicity of the conduction band of the GaN
host. Simply extrapolating our closely spaced data linearly to
vanishing $n_{2D}$, we arrive at $m_0^*$ =0.214$m_e$. This value
is about $\sim$8$\%$ lower than previously published CR data
\cite{Knap96,Knap97}. Since our data contain a small error bar and
extend to very low $n_{2D}$, such a simple extrapolation should
already be quite reliable.

Other groups have previously addressed NP in GaN. For example,
Knap \textit{et al.} explored NP using CR with different $n_{2D}$
(see our Fig.\ref{Fig.2}) and accounted for the magnitude of NP
using a simple two-band approximation \cite{Knap96,Knap97}. In
such an approximation, which includes only coupling between the
lowest conduction and highest valence bands, the effective mass
varies as
\begin{equation}
    \label{eq.1} m^* (E) = m_0^* (1+2 K E/E_g )
\end{equation}
with $K$=1 and $E_g$=3.5 eV. In a 2DES the energy, $E=E_k+ E_F$,
above the band minimum is composed of the average kinetic energy,
$E_k$, of the electrons in the confining potential well and $E_F=
\pi\hbar^2 n_{2D}/m_0^*$ \cite{massNote}, the Fermi energy of the
2D electron system. The value of $E_k$ is dependent on the form of
the wave function of the confined electrons. In a simple
triangular potential approximation, the average kinetic energy is
$E_k = E_c/3$, with $E_c$ being the confinement energy of the
lowest subband \cite{AndoFowlerStern}. Using the more accurate
Fang-Howard variational wave function gives $E_k = \hbar^2 b^2 / 8
m_0^*$, where $b^3=48\pi m_0^* e^2(N_{dep}+\frac{11}{32} n_{2D}) /
\epsilon \hbar^2$ \cite{AndoFowlerStern}. In our samples, since
both the MBE and the HVPE GaN are \textit{n}-type, the depletion
layer density, $N_{dep}$, can be set to zero. We use this two-band
model with only one adjustable parameter, $m_0^*$, to fit the
high-density data of Ref. \cite{Knap97} (dashed line in
Fig.\ref{Fig.2}). $E_k$ was computed in the triangular
approximation following the authors of Ref. \cite{Knap97}.
Clearly, although this procedure can describe the original density
dependence of $m$* of Ref. \cite{Knap97} due to the rather large
error bars, it fails to account for our data. Neither a simple
vertical shift of the line (a different $m_0^*$) nor the usage of
the Fang-Howard wavefunction can resolve this discrepancy. What is
required, is a much stronger dependence of $m$* on $n_{2D}$.

Empirically, we pursue an approach taken by Singleton \textit{et
al.} \cite{Singleton88}, who used a \emph{modified} two-band model
to describe their NP data in GaAs. The authors considered $K$ in
Eq.\ref{eq.1} to be a second fitting parameter. The inclusion of a
variable $K >1$ into the analytic expression incorporates the
influence of higher conduction bands, simulating the results of a
more elaborate, multi-band \textbf{k.p} calculation
\cite{Zawadzki90}. Working with the Fang-Howard model to determine
$E_k$ we find a very good fit to our data for $K$ = 2.5 and $m_0^*
= (0.208\pm0.002)m_e$. Even if we use the less reliable triangular
approximation, the required $K$ = 1.9. The effectiveness of the
modified expression in fitting the density dependence of $m$* over
a wide range of $n_{2D}$ demonstrates that the conduction band of
GaN is more non-parabolic than was previously assumed
\cite{Lin98,Knap96,Knap97,Hang01}.

Before concluding that the proposed NP model appropriately
describes our CR data, we need to assure ourselves that polaron
effects are a negligible contributor to $m$*. This mass
enhancement factor results from electron-LO phonon coupling in
polar semiconductors, such as GaN. A determination of a polaron
mass enhancement in heterostructures requires inclusion of
screening in 2D and the finite width of the electronic wave
function \cite{PauliNote}, since both greatly reduce interaction
between 2D carriers and LO phonons \cite{DasSarma83}. Following
Ref. \cite{DasSarma83}, we calculated the polaron effective mass
in AlGaN/GaN heterostructures using a a Frohlich constant,
$\alpha$ = 0.49 \cite{Drechsler95}, Fang-Howard variational wave
function, and a static Thomas-Fermi screening model. We find a
polaron enhancement of less than 1$\%$ for $m$* for
$n_{2D}=9\times10^{12}$cm$^{-2}$. Since the effect decreases with
decreasing density, corrections for lower density specimens are
smaller yet. This mass enhancement is considerably smaller than
the 10$\%$ estimated previously for $n_{2D}= 3.1\times10^{12}
$cm$^{-2}$ \cite{Knap96}, where screening and finite width had
been neglected. A 1$\%$ mass enhancement due to polaronic coupling
lies within the error bars of our CR data.

The large spread in the value of the band edge mass $m_0^*$ in the
literature is mostly due to two reasons: the underestimation of
the NP and the overestimation of polaronic corrections. Applying
our NP and polaron corrections to the available effective mass
data (Ref.
\cite{Drechsler95,Knap96,Knap97,Hang01,Lin98,Perlin96,Moore95,Witowskia99,
Meyer95,Wang96}), we reach a much more coherent picture for the
band edge effective mass, $m_0^*$, in GaN, as shown in the inset
to Fig. \ref{Fig.2}. We observe that the majority of the values
for $m_0^*$ are very close to an average $m_0^*$ = 0.204$m_e$.
However, most of the $m_0^*$ data from SdH (displayed in stars)
remain at variance from the CR, infra-red reflectivity and donor
spectroscopy data, a fact that remains unexplained.

\begin{figure}
\includegraphics{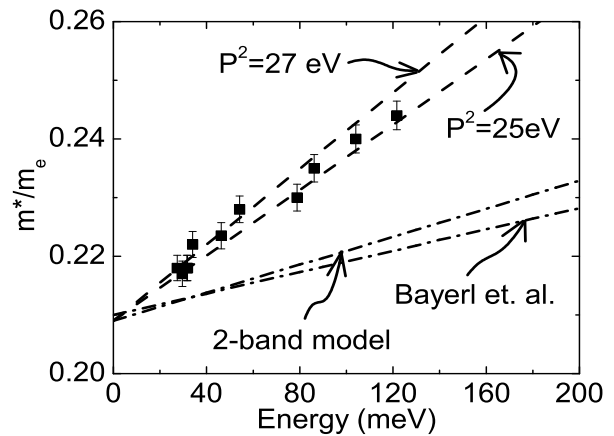}
\caption{\label{Fig.3} Variation of $m$* with energy according to
\textbf{k.p} calclations, assuming $m_0^*$=0.208$m_e$. The
two-band model and the five-band \textbf{k.p} results of Bayerl et
al., which neglect remote bands both underestimate the
energy-dependence of $m$*. The experimental data (solid squares)
can be explained by multi-band \textbf{k.p} calculations that
include the influence of remote bands (dashed lines).}
\end{figure}
Since $m$*$(E)$ cannot be accurately represented with a two-band
model, a five-band \textbf{k.p} calculation \cite{Zawadzki90} in
the zincblende approximation was performed to model our data.
Figure \ref{Fig.3} shows the results. The material parameters
employed using the Koster notation are the spin-orbit splitting,
$\Delta^\prime_0$, of the $\Gamma_5$ conduction band and the
momentum matrix elements, $P^2$ and $\lambda^2P^2$, coupling the
$\Gamma_1$ conduction band with the $\Gamma_5$ valence and
conduction bands. They are not precisely known. The formalism
proposed by Carlos \textit{et al.} \cite{Carlos93} and used by
Bayerl \textit{et al.} \cite{Bayerl01} in a 5-band model to relate
the parameters was utilized to calculate $m$*$(E)$. However, as
seen in Fig.\ref{Fig.3}, the parameters chosen by Bayerl
\textit{et al.} lead to a non-parabolicity even lower than in a
two-level model. We find that the experimental data points can
only be matched if an extra parameter $C$, taking remote bands
into account at $k$=0, is included \cite{Hermann77}. If one
assumes $C$= -1.5 (compare with $C$=-2 for GaAs
\cite{Zawadzki90}), $P^2$ is in the range 25-27 eV with
$\lambda^2\approx$0.33-0.48 and $\Delta^\prime_0\approx$120-180
meV. This set of parameters is close to the ``best'' set of values
chosen by Kennedy \textit{et al.} \cite{Kennedy99} and confirm
that $P^2$ is relatively large for GaN ($P^2\approx$26 eV).

In conclusion, our CR experiments set the conduction band edge
mass in GaN to $m_0^* = (0.208\pm0.002)m_e$. Using our
determination of the nonparabolicity and reviewing the polaron
correction we reach much better agreement between several
published data for $m_0^*$.

\begin{acknowledgments}
We thank M.S. Brandt, L.N. Pfeiffer, K.W. West, and W.I. Wang for
helpful discussions. A portion of the work was performed at the
National High Magnetic Field Laboratory, which is supported by NSF
Cooperative Agreement No.DMR-0084173 and by the State of Florida.
Financial support from the W. M. Keck Foundation is gratefully
acknowledged.
\end{acknowledgments}


\begin{thebibliography}{22}
\bibitem[{mas()}]{massNote}
\bibinfo{note}{Here we use an average mass of $m_0^*$ =0.23$m_e$, since it
  enters $E_F$ as $(m_0^*)^{-1}$ and $E_k$ as $(m_0^*)^ {-1/3}$ and therefore
  enters $m$* in higher order.}
\bibitem[{Pau()}]{PauliNote}
\bibinfo{note}{According to Ref. \cite{DasSarma83} phase-space restrictions due
  to the degeneracy of the 2DES can be ignored for $k_F/q_0$ at least as high
  as 0.6, where $q_0$ is the phonon wavevector and $k_F$ the Fermi wavevector.
  In our AlGaN/GaN systems $k_F/q_0$ $\leq$0.6 for $n_{2D}$
  $\leq$3.4$\times10^{12}$ cm$^{-2}$ and still $k_F/q_0$ $\approx$ 0.8 for our
  highest density. We therefore also ignore phase space restriction.}
\end{thebibliography}

\end{document}